\documentclass{PoS}

\usepackage{geometry}
\usepackage{graphicx}
\usepackage{epsfig}
\usepackage{latexsym}
\usepackage{amsmath}
\usepackage{amsfonts}
\usepackage{amsxtra}
\usepackage{axodraw}

\def\krto{ {\,\,\lower .8ex\hbox {$\longrightarrow \atop k \rightarrow 0$}\,\,}}

\def\bea{\begin{eqnarray} }
\def\beq{\begin{eqnarray} }

\def\eea{\end{eqnarray}}
\def\eeq{\end{eqnarray}}
\def\eq#1{eq.~(\ref{#1})}

\newcommand{\ghostSD}{\begin{picture}(150,25)(0,0)
\SetWidth{1.2}
\DashArrowLine(12.5,0)(37.5,0){5}
\DashArrowLine(37.5,0)(75,0){5}
\DashLine(75,0)(112.5,0){5}
\DashArrowLine(112.5,0)(137.5,0){5}
\SetWidth{1}
\Vertex(112.5,0){2}
\GlueArc(75,0)(37.5,0,90){-4}{6}
\GlueArc(75,0)(37.5,90,180){-4}{6}
\CCirc(75,0){5}{Black}{Yellow}
\CCirc(75,37.5){5}{Black}{Yellow}
\CCirc(37.5,0){5}{Black}{Yellow}
\Text(20,-10)[l]{a,k}
\Text(50,15)[l]{d,$\nu$}
\Text(100,-10)[l]{e}
\Text(100,15)[r]{f,$\mu$}
\Text(50,-10)[l]{c,q}
\Text(120,-10)[l]{b,k}
\Text(75,48)[c]{q-k}
\end{picture}}

\newcommand{\ghostDr}{\begin{picture}(100,25)(0,0)
\SetWidth{1.2}
\DashArrowLine(12.5,0)(50,0){5}
\DashArrowLine(50,0)(87.5,0){5}
\CCirc(50,0){5}{Black}{Yellow}
\Text(12.5,-10)[l]{a}
\Text(87.5,-10)[r]{b}
\Text(50,-10)[c]{k}
\end{picture}}

\newcommand{\ghostBr}{\begin{picture}(100,25)(0,0)
\SetWidth{1.2}
\DashArrowLine(12.5,0)(87.5,0){5}
\Text(12.5,-10)[l]{a}
\Text(87.5,-10)[r]{b}
\Text(50,-10)[c]{k}
\end{picture}}

\newcommand{\gluonSDi}{\begin{picture}(112.5,18.75)(0,0)
\SetScale{0.75}
\SetWidth{1.2}
\Gluon(12.5,0)(37.5,0){-4}{2}
\Gluon(37.5,0)(75,0){-4}{3}
\Gluon(75,0)(112.5,0){-4}{3}
\Gluon(112.5,0)(137.5,0){-4}{2}
\SetWidth{1}
\Vertex(112.5,0){2}
\GlueArc(75,0)(37.5,0,90){-4}{6}
\GlueArc(75,0)(37.5,90,180){-4}{6}
\CCirc(75,0){5}{Black}{Yellow}
\CCirc(75,37.5){5}{Black}{Yellow}
\CCirc(37.5,0){5}{Black}{Yellow}
\end{picture}}

\newcommand{\gluonSDii}{\begin{picture}(112.5,18.75)(0,0)
\SetScale{0.75}
\SetWidth{1.2}
\Gluon(15,-5)(75,-5){-3}{4}
\Gluon(75,-5)(135,-5){-3}{4}
\SetWidth{1}
\GlueArc(75,18.75)(18.75,-90,90){3}{4}
\GlueArc(75,18.75)(18.75,90,270){3}{4}
\CCirc(75,-2.5){5}{Black}{Yellow}
\CCirc(75,37.5){5}{Black}{Yellow}
\end{picture}}

\newcommand{\gluonSDiii}{\begin{picture}(112.5,18.75)(0,0)
\SetScale{0.75}
\SetWidth{1.2}
\Gluon(12.5,0)(37.5,0){-4}{2}
\DashLine(37.5,0)(75,0){4}
\DashLine(75,0)(112.5,0){4}
\Gluon(112.5,0)(137.5,0){-4}{2}
\SetWidth{1}
\Vertex(112.5,0){2}
\DashCArc(75,0)(37.5,0,90){4}
\DashCArc(75,0)(37.5,90,180){4}
\CCirc(75,3){5}{Black}{Yellow}
\CCirc(75,37.5){5}{Black}{Yellow}
\CCirc(37.5,3){5}{Black}{Yellow}
\end{picture}}

\newcommand{\gluonSDiv}{\begin{picture}(112.5,18.75)(0,0)
\SetScale{0.75}
\SetWidth{1.2}
\Gluon(15,-5)(75,-5){-3}{4}
\Gluon(75,-5)(135,-5){-3}{4}
\SetWidth{1}
\DashCArc(75,18.75)(18.75,-90,90){4}
\DashCArc(75,18.75)(18.75,90,270){4}
\CCirc(75,-2.5){5}{Black}{Yellow}
\CCirc(75,37.5){5}{Black}{Yellow}
\end{picture}}

\title{The low-momentum ghost dressing function and the gluon mass}

\ShortTitle{The low-momentum ghost dressing...} 

\author{\speaker{J. Rodr\'{\i}guez-Quintero} 
\\
Dpto. F\'isica Aplicada, Fac. Ciencias Experimentales; 
Universidad de Huelva, 21071 Huelva; Spain.\\
        E-mail: \email{jose.rodriguez@dfaie.uhu.es}}

\abstract{We study both regular (the zero-momentum 
ghost dressing function not diverging), also named decoupling, and 
critical (diverging), also named scaling, Yang-Mills propagators solutions
by analyzing the low-momentum behaviour of the ghost propagator Dyson-Schwinger 
equation (DSE) in Landau gauge, assuming for the truncation a constant ghost-gluon vertex, 
as it is extensively done, and a simple model for a massive gluon propagator. 
The asymptotic expression obtained for the regular or decoupling ghost dressing function 
up to the order ${\cal O}(q^2)$ fits pretty well 
the low-momentum ghost propagator obtained through the numerical integration of 
the coupled gluon and ghost DSE in the PT-BFM scheme and, when the size of the 
coupling renormalized at some scale approaches some critical value, the PT-BFM results
seems to trend to the the scaling solution as a limiting case. }

\FullConference{Light Cone 2010 - LC2010\\
		June 14-18, 2010\\
		Valencia, Spain}

\begin{document}

\section{Introduction}

The low-momentum behaviour of the Yang-Mills propagators derived either from 
the tower of Dyson-Schwinger equations (DSE) or from Lattice simulations in 
Landau gauge has been a very interesting and hot topic for the 
last few years. It seems by now well established that, if we assume 
in the vanishing momentum limit a ghost dressing function behaving as 
$F(q^2) \sim (q^2)^{\alpha_F}$ and a gluon propagator as 
$\Delta(q^2) \sim (q^2)^{\alpha_G-1}$ (or, by following a notation commonly used,
a gluon dressing function as $G(q^2)= q^2 \Delta(q^2) \sim (q^2)^{\alpha_G}$), 
two classes of solutions may emerge (see, for instance, the discussion 
of refs.~\cite{Boucaud:2008ji,Boucaud:2008ky}) from the DSE:
(i) those, dubbed {\it ``decoupling''}, where $\alpha_F=0$ and the suppression of 
the ghost contribution to the gluon propagator DSE results in a massive gluon 
propagator (see \cite{Aguilar:2006gr,Aguilar:2008xm} and references therein); 
and (ii) those, dubbed {\it ``scaling''}, where $\alpha_F \neq 0$ 
and the low-momentum behaviour of both gluon and ghost propagators 
are related by the coupled system of DSE through the condition $2 \alpha_F+\alpha_G = 0$ 
implying that $F^2(q^2)G(q^2)$ goes to a non-vanishing constant when $q^2 \to 0$ 
(see \cite{Alkofer:2000wg,Fischer:2008uz} and references therein). 

Lattice QCD results appear to support only the massive gluon ($\alpha_G=1$) or scaling solutions  
(see~\cite{Cucchieri:2007md,Bogolubsky:2007ud,IlgenGrib,Boucaud:2005ce,Oliveira:2010xc,Bornyakov:2009ug} 
and references therein), and also pinching technique results (see, for instance, 
\cite{Cornwall,Binosi:2002ft} and references therein), 
refined Gribov-Zwanziger 
formalism (see~\cite{Dudal:2007cw}) or other approaches like the infrared mapping of $\lambda \phi^4$ and Yang-Mills 
theories in ref.~\cite{Frasca:2007uz} or the massive extension of the Fadeev-Popov 
action in ref.~\cite{Tissier:2010ts} appear to point to.

In the present note, we briefly review the work of refs.~\cite{Boucaud:2008ji,Boucaud:2008ky,Boucaud:2010gr}, where  
it is established how both types of IR solutions for Landau gauge DSE emerge and how the transition between them may occur,
and that of ref.~\cite{RodriguezQuintero:2010xx} which extends the previous studies by the analysis of the 
results~\cite{Aguilar} obtained by solving the coupled system of Landau gauge ghost and gluon propagators DSE 
within the framework of the pinching technique in the background field
 method~\cite{Binosi:2002ft} (PT-BFM)

\section{The two kinds of solutions of the ghost propagator Dyson-Schwinger equation}\label{revisiting}
\label{twosol}

As was explained in detail in refs.~\cite{Boucaud:2008ky,Boucaud:2010gr,RodriguezQuintero:2010xx}, the low-momentum behavior 
for the Landau gauge ghost dressing function can be inferred from the analysis
of the Dyson-Schwinger equation for the ghost
propagator (GPDSE),  which can be written diagrammatically as

\vspace{\baselineskip}
\begin{small}
\bea
\left(\ghostDr\right)^{-1}%
=
\left(\ghostBr\right)^{-1}%
- 
\ghostSD \ , 
\label{ghostSD}
\eea\end{small}%

\noindent That analysis is performed on a very general ground: one applies the MOM renormalization 
prescription,
\beq
F_R(\mu^2) \ = \ \mu^2 \Delta_R(\mu^2) \ = \ 1 \ ,
\eeq
where $\mu^2$ is the subtraction point, chooses for the ghost-gluon vertex, 
\beq
\widetilde{\Gamma}_\nu^{abc}(-q,k;q-k) \ = \
i g_0 f^{abc} \left( \ q_\nu H_1(q,k) + (q-k)_\nu H_2(q,k) \ \right) \ ,
\label{DefH12}
\eeq
to apply this MOM prescription in Taylor kinematics 
({\it i.e.} with a vanishing incoming ghost momentum) 
and assumes the non-renormalizable bare ghost-gluon form factor, $H_1(q,k)=H_1$, 
to be constant in the low-momentum regime for the incoming ghost.
Then, the low momentum-behaviour of the ghost dressing function and the gluon propagator is 
supposed to be well described by 
\beq\label{gluonprop}
\Delta_R(q^2) &=& \frac{B(\mu^2)}{q^2 + M^2} \ 
\simeq \frac{B(\mu^2)}{M^2} \left( 1 - \frac{q^2}{M^2} + \cdots \right) \ ,
\\ \label{dress}
F_R(q^2) &=& A(\mu^2) \left( \frac{q^2}{M^2} \right)^{\alpha_F} \left( 1 + \cdots 
\rule[0cm]{0cm}{0.6cm} \right) \ ,
\eeq
and one is finally left with:
\beq \label{solsFs}
F_R(q^2) \simeq 
\left\{
\begin{array}{lr}
\displaystyle 
\left(
\frac {10 \pi^2}{N_C H_1 g_R(\mu^2) B(\mu^2)} 
\right)^{1/2}
\ \left(\frac {M^2} {q^2} \right)^{1/2} \ 
\left( 1 + \cdots 
\rule[0cm]{0cm}{0.6cm} \right)
&
\mbox{\rm if} \ \alpha_F \neq 0
\\
\displaystyle 
F_R(0) \left( 1 \ + \ 
\frac{N_C H_1}{16 \pi} \ \overline{\alpha}_T(0) \ 
\frac{q^2}{M^2} \left[ \ln{\frac{q^2}{M^2}} - \frac {11} 6 \right]
\ + \ {\cal O}\left(\frac{q^4}{M^4} \right) \right) 
\rule[0cm]{0.75cm}{0cm}
\rule[0.5cm]{0cm}{0.5cm}
& 
\mbox{\rm if} \ \alpha_F = 0
\end{array}
\right.
\eeq
where 
\beq\label{coefC2}
\overline{\alpha}_T(0) =  M^2 \frac{g^2_R(\mu^2)}{4 \pi} 
F_R^2(0) \Delta_R(0) .
\eeq
It should be understood that the subtraction momentum for all the renormalization quantities is $\mu^2$. 
The case $\alpha_F \neq 0$ leads to the so-called scaling solution, where the low-momentum behavior of 
the massive gluon propagator forces the ghost dressing function to diverge at low-momentum 
through the scaling condition: $2 \alpha_F + \alpha_G=0$ ($\alpha_G=1$ is the power 
exponent when dealing with a massive gluon propagator). As this scaling condition is verified, 
the perturbative strong coupling defined in this Taylor scheme~\cite{Boucaud:2008gn}, 
$\alpha_T=g_T^2/(4\pi)$, has to reach a constant at zero-momentum,
\beq
\alpha_T(0) \ = \frac{g^2(\mu^2)}{4 \pi} \lim_{q^2\to 0} q^2 \Delta(q^2) F(q^2) F^2(q^2) \ ,
\ = \ \frac{5 \pi}{2 N_C H_1}
\eeq
as can be obtained from Eqs.(\ref{gluonprop},\ref{solsFs}). 
The case $\alpha_F=0$ corresponds to the so-called decoupling solution, 
where the zero-momentum ghost dressing function reaches a non-zero finite value 
and \eq{solsFs} provides us with the first asymptotic corrections to this 
leading constant. This subleading correction is controlled by the zero-momentum value of 
the coupling defined in \eq{coefC2}, which is an extension of the non-perturbative effective 
charge definition from the gluon propagator~\cite{Aguilar:2008fh} to the Taylor 
ghost-gluon coupling~\cite{Aguilar:2009nf}. 
As a consequence of the appropriate {\it amputation} of a 
massive gluon propagator, where the gluon mass scale is the same RI-invariant 
mass scale appearing in \eq{gluonprop}, 
this Taylor effective charge is frozen at low-momentum and gives 
a non-vanishing zero-momentum value.

\section{Comparison with numerical results from coupled DSE's}
\label{comparing}

We shall now compare the formulas given by eqs.~(\ref{gluonprop},\ref{solsFs}) 
with some numerical results for the gluon propagator 
and ghost dressing function obtained by solving the coupled system
of gluon and ghost DS equations obtained by applying the pinching technique 
in the background field method (PT-BFM)~\cite{Binosi:2002ft} 
(see also \cite{Binosi:2009qm} and references therein). 
In the PT-BFM scheme for the coupled DSE system, the ghost propagator DSE 
is the same of Eq.~(\ref{ghostSD}), where the approximation $H_1=1$, and 
the gluon DSE is given by 
\beq\label{coupledDSE}
\frac{(1+G(q^2))^2}{\Delta(q^2)} \left( g_{\mu\nu} - \frac{q_\mu q_\nu}{q^2} \right) =   
q^2 g_{\mu\nu} - q_\mu q_\nu + i \sum_{i=1}^4 \left( a_i \right)_{\mu\nu}
\eeq
where
\beq\label{gluondiags}
a_1 = \gluonSDi ,  & a_2 = \gluonSDii 
\nonumber \\
a_3 = \rule[0cm]{0cm}{1.5cm} \gluonSDiii , & a_4= \gluonSDiv .
\eeq
where the external gluons are treated, from the point of view of Feynman rules, as background fields 
(these diagrams should be also properly regularized, as explained in \cite{Binosi:2009qm}). 
The function $1+G$ defined in ref.~\cite{Grassi:1999tp} can be, 
in virtue of the ghost propagator DSE, connected to the ghost propagator~\cite{Aguilar:2009nf}. 
The coupled system is to be solved, by numerical integration, 
with the two following boundary conditions as the only required inputs:
the zero-momentum value of the gluon propagator and that of the coupling at 
a given perturbative momentum, $\mu=10$ GeV, that will be used as 
the renormalization point.

Thus, The PT-BFM framework leaves us with an attractive model for gluon 
and ghost propagators providing quantitative description of lattice 
data~\cite{Aguilar:2008xm,Aguilar:2010gm} and giving well account of their main qualitative features: 
finite gluon propagator and finite ghost dressing function at zero-momentum. 
Futhermore, the coupled DSE system can be solved with different boundary 
conditions (see below).In particular, solutions obtained by keeping the zero-momentum value of the 
gluon propagator fixed (see lefthand plots of fig.~\ref{fig:ghgl}) while $\alpha(\mu^2=100\mbox{\rm~GeV}^2)$
 is ranging from 0.15 to 0.1817 are available~\cite{Aguilar}. 
these solutions can be confronted to the asymptotical 
expressions derived in the previous section.


\subsection{Decoupling solutions}

\begin{figure}[hbt!]
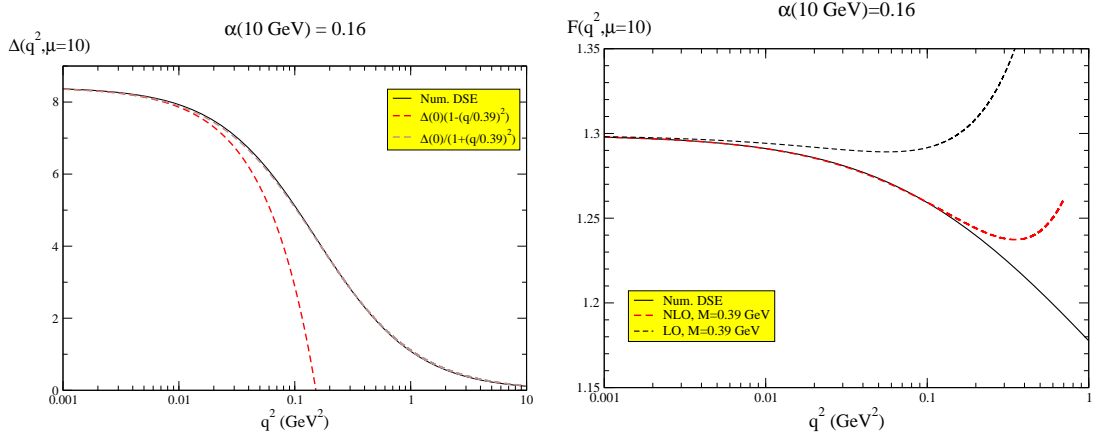

\begin{center}
\begin{tabular}{cc}
\includegraphics[width=7cm]{FIGS/gluon016.eps} &
\includegraphics[width=7cm]{FIGS/ghost016-newF.eps} \\
\end{tabular}
\end{center}
\caption{\small Gluon propagator (left) and ghost dressing function (right) after the numerical integration of  
the coupled DSE system for $\alpha(\mu=10 \mbox{\rm GeV})=0.16$ taken from \cite{Aguilar} . The curves 
for the best fits to gluon propagator and ghost propagator data explained in the text appear as red dotted lines. 
the same for the black dotted line in the lefthand plot but retaining only the logarithmic leading term for 
the asymptotic ghost dressing function by dropping the $-11/6$ away. In the righthand plots, the red dotted 
lines correspond to apply.}
\label{fig:ghgl}
\end{figure}

Then, as the gluon propagator solutions in the PT-BFM scheme result to behave as massive ones, 
the eqs.~(\ref{gluonprop},\ref{solsFs}) must account for the low-momentum behaviour of 
both gluon propagator and ghost dressing function with $H_1=1$ and 
$\overline{\alpha}_T(0)$ given by \eq{coefC2}, with $\alpha_T(\mu^2)=g_R^2(\mu^2)/(4\pi)$ being fixed, 
as a boundary condition for the numerical integration of the coupled DSE for each particular 
solution of the family (see tab. \ref{tab-fits}). Furthermore, the zero-momentum values of the 
ghost dressing function, $F_R(0)$ and of the 
gluon propagator, $\Delta_R(0)$, can be taken from the numerical solutions of the DSE (for any value of the 
$\alpha(\mu=10 \mbox{\rm GeV})$). These altoghether with the gluon masses obtained by the fit of 
\eq{gluonprop} to the numerical DSE gluon propatator solutions  (see 
the left plot in fig.~\ref{fig:ghgl}, for $\alpha(\mu)=0.16$, and the results for $\alpha(\mu)=0.15, 0.16, 0.17$ in 
tab.~\ref{tab-fits}, taken from ref.~\cite{RodriguezQuintero:2010xx}), provide us with all the ingredients 
to evaluate, with no unknown parameter, \eq{solsFs}. 

Indeed, the expression given by \eq{solsFs} can be succesfully
applied to describe the solutions all over the range of coupling values, 
$\alpha(\mu)$, at $\mu=10$ GeV (provided that they are not very close of the critical 
coupling that will be defined in the next subsection). 
This can be seen, for instance, for $\alpha=0.16$, 
in the right plots of fig.~\ref{fig:ghgl} and it is shown for $\alpha=0.15,0.16,0.17$ en 
ref.~\cite{RodriguezQuintero:2010xx}.

\begin{table}
\begin{center}
\begin{tabular}{|c||c|c|} 
\hline
\rule[0cm]{0cm}{0.5cm} $\alpha(\mu)$ & $\overline{\alpha}_T(0)$ & $M$ (GeV) [gluon] \\
\hline
\hline
0.15 & 0.24 & 0.37 
\\
\hline
0.16 & 0.30 & 0.39 \\
\hline
0.17 & 0.41 & 0.43 \\
\hline
\end{tabular}
\end{center}
\caption{\small Gluon masses and the zero-momentum non-perturbative effective 
charges, taken from ref.~\cite{RodriguezQuintero:2010xx} and obtained as discussed in 
the text.
}
\label{tab-fits}
\end{table}

\subsection{The ``critical'' limit}

\begin{figure}[!hbt]
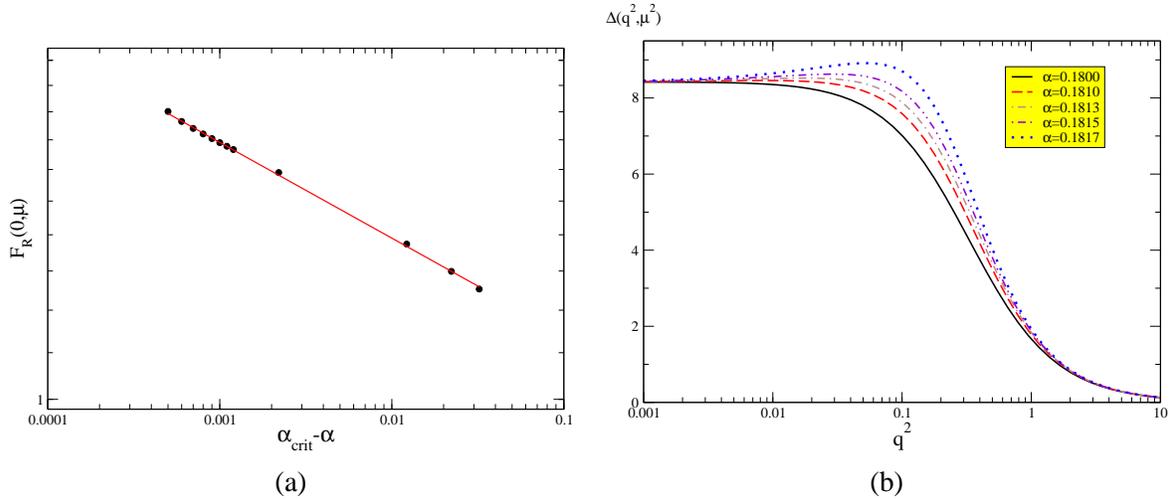

\begin{center}
\begin{tabular}{cc}
\includegraphics[width=7.5cm]{FIGS/ghost0-F-0.1822.eps} 
&
\includegraphics[width=7.5cm]{FIGS/CritGluons.eps} 
\\
(a) & (b) 
\end{tabular}
\end{center}
\caption{\small (a) Log-log plot of the zero-momentum values of the ghost dressing function, obtained 
by the numerical integration of the coupled DSE system in the PT-BFM scheme, in terms of 
$\alpha_{\rm crit}-\alpha$. $\alpha=\alpha(\mu=10 \mbox{\rm GeV})$, 
the value of the coupling at the renormalization momentum, is an initial condition for 
the integration; while $\alpha_{\rm crit}$ is fixed to be 0.1822, as explained in the text, 
by requiring the best linear correlation. 
(b) Gluon propagator solutions
in terms of $q^2$ for the same coupled DSE system for different values of $\alpha(\mu=10 \mbox{\rm GeV})$, 
all very close to the critical value, ranging from 0.18 to 0.1817~.}
\label{fig:ghost0s}
\end{figure}

There appears to be a {\it critical} value of the coupling, 
$\alpha_{\rm crit}=\alpha(\mu^2)\simeq 0.182$ with $\mu=10\mbox{\rm ~Gev}$, 
above which the coupled DSE system does not converge any longer to a solution~\cite{Aguilar}.
As a matter of the fact, we know from refs.\cite{Boucaud:2008ky,RodriguezQuintero:2010xx} that
the scaling solution implies for the coupling
\beq\label{ap:crit}
\alpha_{\rm crit} \ = \ \frac{g_R^2(\mu^2)}{4 \pi} 
\simeq
\frac{5 \pi^2}{2 N_C A^2(\mu^2) B(\mu^2) } \ ,
\eeq
where $B(\mu^2)$ is determined by the gluon propagator solution that is supposed to 
behave as \eq{gluonprop}, and $A(\mu^2)$ by the ghost propagator that should behave 
as \eq{solsFs} in the case $\alpha_F \neq 0$.
Again, $\mu^2$ is the momentum at the subtraction point.
This is also shown in ref.~\cite{Boucaud:2008ji}, where only the ghost propagator 
DSE with the kernel for the gluon loop integral is obtained from gluon propagator lattice 
data. In the analysis of ref.~\cite{Boucaud:2008ji}, a ghost dressing 
function solution diverging at vanishing momentum appears to exist and 
verifies eqs.~(\ref{solsFs},\ref{ap:crit}), while regular or decoupling solutions
 exist for any $\alpha < \alpha_{\rm crit}$. In ref.~\cite{RodriguezQuintero:2010xx}, 
 a more complete analysis is performed by studying again the  
dressing function computed by solving \eq{coupledDSE} for  
the different values of the coupling, $\alpha=\alpha(\mu^2)$, at $\mu^2=100$ 
GeV$^2$~\cite{Aguilar}. The ghost dressing function at vanishing momentum, $F(0,\mu^2)$, 
is described by the following power behaviour,
\beq
F(0) \ \sim \ (\alpha_{\rm crit} - \alpha(\mu^2))^{- \kappa(\mu^2)} \ ,
\eeq
where $\kappa(\mu^2)$ is a critical exponent (depending presummably on the 
renormalization point, $\mu^2$), supposed to be positive and to govern the transition 
from decoupling ($\alpha < \alpha_{\rm crit}$) to the scaling ($\alpha = \alpha_{\rm crit}$) 
solutions; and where we let $\alpha_{\rm crit}$ be a free parameter to be fitted 
by requiring the best linear correlation for $\log[F(0)]$ in terms of 
$\log[\alpha_{\rm crit}-\alpha]$. In doing so, the best correlation 
coefficient is 0.9997 for $\alpha_{\rm crit}=0.1822$, which is pretty close to 
the critical value of the coupling above which the coupled DSE system does not converge 
any more, and $\kappa(\mu^2) = 0.0854(6)$. This can be seen in fig.~\ref{fig:ghost0s}.(a), where 
the log-log plot of $F_R(0)$ in terms of $\alpha_{\rm crit}-\alpha$ is shown and 
the linear behaviour with negative slope corresponding to the best correlation coefficient
strikingly indicates a zero-momentum ghost propagator diverging 
as $\alpha \to \alpha_{\rm crit}$. Nevertheless, no critical or scaling solution 
appears for the coupled DSE system in the PT-BFM, although the decoupling solutions obtained for any 
$\alpha < \alpha_{\rm crit} = 0.1822$ seem to approach the behaviour of a scaling one 
when $\alpha \to \alpha_{\rm crit}$. 
The absence of the scaling solution in the PT-BFM approach can be well understood by 
analysing \eq{coupledDSE} as explained in ref.~\cite{RodriguezQuintero:2010xx}.

\vspace{0.9cm}

When approaching the critical value of the coupling, the gluon propagators obtained 
from the coupled DSE system in PT-BFM must be also thought to obey the 
same critical behaviour pattern as the ghost propagator. 
In the PT-BFM, the value at zero-momentum being 
fixed by construction~\cite{Aguilar:2008xm,Aguilar}, one should expect that, 
instead of decreasing, the 
gluon propagator obtained for couplings near to the critical value increases for low momenta: 
the more one approaches the critical coupling the more it has to increase. 
This is indeed the case, as can be seen in fig.~\ref{fig:ghost0s}(b). 
This also implies that, near the critical value, the low momentum propagator does not 
obey \eq{gluonprop} and that consequently \eq{solsFs} does not work any longer to 
describe the low momentum ghost propagator.

\section{Conclusions}\label{conclu}

The ghost propagator DSE, with the only assumption of taking $H_1(q,k)$ from the 
ghost-gluon vertex in \eq{DefH12} to be constant in the infrared domain of $q$, can be 
exploited to look into the low-momentum behaviour of the ghost propagator.  The
two classes of solutions named ``decoupling'' and ``scaling'' can be indentified and 
shown to depend on whether the ghost dressing function achieves a finite non-zero
constant ($\alpha_F=0$) at vanishing momentum or not ($\alpha_F \neq 0$). The 
solutions appear to be dialed by the size of the coupling at the renormalization 
momentum which plays the role of a boundary condition for the DSE integration. 

We applied a model with a massive gluon propagator to obtain the low-momentum 
behaviour of the ghost propagator that results to be regulated by the 
gluon mass and by a regularization-independent 
dimensionless quantity that appears to be the effective charge defined from 
the Taylor-scheme ghost-gluon vertex at zero momentum.
Then, we demonstrated that the asymptotic decoupling formula ($\alpha_F=0$) 
successfully describes the low-momentum ghost propagator computed trhough 
the numerical integration of the coupled gluon and ghost DSE in the PT-BFM scheme. 
We also show that the zero-momentum ghost dressing function 
tends to diverge when the value of the coupling dialing the solutions 
approaches some critical value. Such a divergent behaviour at the critical coupling 
corresponds to a scaling solution where, if the gluon 
is massive, $\alpha_F=-1/2$. 

\bigskip

{\bf Acknowledgements:} 
The author acknowledges the Spanish MICINN for the 
support by the research project FPA2009-10773 and ``Junta de Andalucia'' by P07FQM02962.

\end{document}